\RequirePackage{fix-cm}
\documentclass[twocolumn,epjc3]{svjour3}
\smartqed  
\RequirePackage{graphicx}
\RequirePackage[numbers,sort&compress]{natbib}
\usepackage{amsmath}
\usepackage{amssymb}
\usepackage{latexsym}
\usepackage{graphicx}
\journalname{Eur. Phys. J. C}
\begin{document}

\title{Quasinormal Modes and  Thermodynamics  of  Linearly Charged BTZ Black holes in  Massive Gravity  in (Anti)de Sitter Space Time
} 

\titlerunning{QNMs and Thermodynamics of linearly Charged BTZ Black Holes in Massive Gravity}        

\author{Prasia P\thanksref{e1}
        \and
        Kuriakose V C\thanksref{e2} 
}

\thankstext{e1}{e-mail: prasiapankunni@cusat.ac.in}
\thankstext{e2}{e-mail: vck@cusat.ac.in}

\institute{Department  of  Physics, Cochin University of Science and
Technology, Kochi-682022, India\label{e1,e2}
          }

\date{Received: date / Accepted: date}

\maketitle

\begin{abstract}
In this work we study the Quasi Normal Modes(QNMs) under massless
scalar perturbations and the thermodynamics of linearly charged BTZ
black holes in  massive gravity in the (Anti)de Sitter((A)dS) space
time. It is found that the behavior of QNMs changes with the massive
parameter and also with the charge of the black hole. The
thermodynamics of such black holes in the (A)dS space time is also
analyzed in detail. The behavior of specific heat with temperature
for such black holes gives an indication of a phase transition that
depends on the massive parameter and also on the charge of the black
hole.
\end{abstract}
\section{Introduction}

Einstein's General Theory of Relativity(GTR) helped us to understand
the dynamics of the universe. But there are some fundamental issues
that could  not be addressed in GTR \cite{eref1}  and several
attempts are being made to modify the GTR to  find  solutions to
these fundamental issues. GTR is  a theory  based on massless
gravitons with two degrees of freedom. A  way  of modifying GTR
essentially implies giving mass to the graviton  and  in  the
present  study we consider massive gravity. The attempts to modify
GTR resulted in the so called \lq Alternative Theories of Gravity\rq
\cite{1}. Theories concerning the breaking up of Lorentz invariance
and spin had been explored in depth\cite{2}. The first attempt
towards constructing a theory of  massive gravity was done by Fierz
and Pauli\cite{3} in 1939. Only  by 1970s researchers showed
interests in this formulation. van Dam and Veltman\cite{4} and
Zhakharov\cite{5} in 1970 showed  that  a theory of  massive gravity
could never resemble GTR in the massless limit and this is known as
vDVZ discontinuity. Later Vainshtein\cite{6} proposed that the
linear massive gravity can be recovered to GTR through the \lq
Vainshtein Mechanism\rq\, at small scales by including non-linear
terms in the hypothetical massive gravity action.  But this model
suffers from a pathology called \lq Boulware-Deser\rq (BD) ghost and
was ruled out on the basis of solar system tests\cite{7}. Later a
class of massive gravity was proposed by de Rham, Gahadadze and
Tolley called \lq dRGT massive gravity\rq\, that evades the BD
ghost\cite{8,9}. In this theory the mass terms were produced by a
reference metric. A class of black hole solutions in the dRGT model
and their thermodynamic behavior were studied  later\cite{10,11,12}.
Vegh\cite{13} proposed another type of massive gravity theory. This
theory was similar to dRGT except that the reference metric was a
singular one. Using this theory he showed that graviton behaves like
a lattice and showed Drude peak. This theory was found to be
ghost-free and stable for arbitrary singular metric. \\ \\
It was Hawking\cite{14} who first showed that black holes thermally
radiate and calculated its temperature. Thereafter the
thermodynamics of black holes got wide acceptance and interests
among researchers. The question of thermal stability is one of the
important aspects of black hole thermodynamics\cite{15,16}. The
thermodynamics  and phase transition shown by black holes have been
largely explored for almost all space times\cite{17,18,19,20} and
references cited therein. In the realm of massive gravity also, the
thermodynamics and phase transitions have been studied for different
black hole space time\cite{21,22}.\\ \\
Recently there has been a growing interest in the asymptotically
Anti de Sitter(AdS) spacetimes. The black hole solution proposed by
Banados-Teitelboim-Zanelli (BTZ) in  $(2+1)$ dimensions deal with
asymptotically AdS space time and has got well defined charges at
infinity, mass, angular momentum and makes a good testing ground
especially when one would like to go beyond the asymptotic
flatness\cite{23}. Another interesting aspect of the black hole
solution is related to the AdS/CFT (Conformal Field Theory)
correspondence. In $(2+1)$ dimensions, the BTZ black hole solution
is a space time of constant negative curvature and it differs from
the AdS space time in its global properties\cite{24}. The
thermodynamic phase transitions and area spectrum of the BTZ black
holes are studied in detail\cite{25,26,27}. Also, the charged BTZ
black hole solutions are studied for the phase transition in
Ref.\cite{28,29}.\\ \\
Another important aspect of a black hole is its Quasi Normal Modes
(QNMs). QNMs can be found out as a solution to the perturbed field
equation corresponding to the scalar, gravitational and
electromagnetic perturbations of black hole space time. It comes out
as a natural response to these perturbations. The existence of QNMs
was first found by Visweshwara\cite{30} and attempts were made to
find out QNMs for different space times. QNMs of black holes were
first numerically computed by Chandrasekhar and Detweiler\cite{31}.
It was Cardoso and Lemos\cite{32} who first calculated the exact
QNMs of the BTZ black holes. They have found out both analytical and
numerical solutions to the BTZ black hole perturbation for
non-rotating BTZ black holes. It is interesting to note that they
got exact analytical solutions to the wave equation that made BTZ an
important space time where one can prove or disprove the conjectures
relating to QNMs, critical phenomena or area quantization.\\ \\
Electromagnetic field can be a good choice of source for getting
deep insights into the 3 dimensional massive gravity. In this paper
the QNMs, the associated phase transition and thermodynamics of  BTZ
black hole in massive  gravity in the presence of Maxwell's field
has been studied. The paper is organized as follows: In section $2$
the QNMs of a linearly charged BTZ black holes in  massive  gravity
are studied for different values of the massive parameter and charge
for de Sitter and Anti de Sitter space-times. The behavior of quasi
normal frequencies and phase transition are also dealt with. Section
$3$ deals with the thermodynamics of such black holes. The influence
of the massive parameter and charge of the black hole on the various
thermodynamic factors are studied. Section $4$ concludes the paper.
\section{Quasi normal modes of a linearly charged BTZ black hole in massive gravity}
In this section, we first look into the perturbation of black hole
space time by a scalar field. For a linearly charged black hole, the
Einstein-Maxwell action in $(2+1)$ dimension is given by\cite{33},
\begin{equation}\label{}
    S_{EM} = \frac{1}{16\pi G}\int d^3 x\sqrt{-g}\left[R + \frac{2}{l^2} - 4\pi G F_{\mu\nu}F^{\mu\nu}\right],
\end{equation}
where $R$ is the Ricci scalar, $F_{\mu\nu}=\partial_{\mu}A_{\nu} -
\partial_{\nu}A_{\mu}$ is the Faraday tensor, $A_{\mu}$ is the gauge
potential, and $F^{\mu\nu}F_{\mu\nu}$ is the Maxwell invariant. The
action given above can be generalized to include the massive gravity
for the de Sitter space time as\cite{34},
\begin{equation}\label{}
\begin{split}
    S &= \frac{-1}{16\pi}\int d^3 x\sqrt{-g}[R + 2\Lambda + L(\mathcal{F})+\\
    &\,\,\,\,\,\, m^2\sum_{i}^{4} c_i \mathcal{U}_i(g,f)],
\end{split}
\end{equation}
where $\mathcal{F}=F^{\mu\nu}F_{\mu\nu}$, $L$ is an arbitrary
Lagrangian of electrodynamics, $\frac{1}{l^2}=\Lambda$, the
cosmological constant in the de Sitter (dS) space time,
$\mathcal{U}_i$ is the effective potential, $m$ is the massive
parameter and $c_i$s are constants. Varying $(2)$ with respect to
the metric $g_{\mu\nu}$, we can obtain the gravitation field
equation as,
\begin{equation}\label{}
    G_{\mu\nu} + \Lambda g_{\mu\nu} - \frac{1}{2}g_{\mu\nu}L(\mathcal{F}) -
    2 L_{\mathcal{F}}F_{\mu\rho}F^{\rho}_{\nu} + m^2 \mathcal{K}_{\mu\nu} =0,
\end{equation}
where,\\ \\
$G_{\mu\nu}=R_{\mu\nu}-\frac{1}{2}g_{\mu\nu}R,$
$L_{\mathcal{F}}=\frac{dL(\mathcal{F})}{d\mathcal{F}}$,
$\mathcal{K}^{\mu}_{\nu}=\sqrt{g^{\mu\alpha}f_{\alpha\nu}},$ \\ \\
and,\\ \\
\begin{equation}\label{}
    \begin{split}
    \mathcal{K}_{\mu\nu}&=\frac{-c_1}{2}(\mathcal{U}g_{\mu\nu} - \mathcal{K}_{\mu\nu}) - \frac{c_2}{2}(\mathcal{U}_2 g_{\mu\nu} -
     2\mathcal{U}_1 \mathcal{K}_{\mu\nu} +\\
     & 2\mathcal{K}_{\mu\nu}^2) - \frac{-c_3}{2}(\mathcal{U}_3 g_{\mu\nu} -
    4\mathcal{U}_3\mathcal{K}_{\mu\nu} +\\
    &12\mathcal{U}_2 \mathcal{K}_{\mu\nu}^2
    - 24\mathcal{U}_1 \mathcal{K}_{\mu\nu}^3 +
    24\mathcal{K}_{\mu\nu}^4).
    \end{split}
\end{equation}
To obtain static charged black hole solution we consider the $3$
dimensional metric,
\begin{equation}\label{}
    ds^2 = -f(r) dt^2 + f^{-1}(r) dr^2 + r^2 d\theta^2.
\end{equation}
To get an exact solution for this metric, the following ansatz is
employed\cite{13},
\begin{equation}\label{}
    f_{\mu\nu} = diag(0,0,c^2 h_{ij}),
\end{equation}
where $c$ is a positive constant. One of the solutions after proper
rescaling leads to the metric function in the dS space
as\cite{33,34},
\begin{equation}\label{}
    f(r) = \Lambda r^2 - m_0 - 2Q\ln{\frac{r}{\alpha}} + m^2 c c_1 r,
\end{equation}
where $m_0$ is related to the mass of the black hole, $Q$ is the
charge parameter, $\alpha$ is an arbitrary constant and $c_1$ is a
constant. For an Anti de Sitter space, $\Lambda$ will take negative
values. From the metric function, it can be understood that the
contribution of the massive term depends on the sign of $c_1$. In
this Section, we look into the behavior of QNMs of the linearly
charged BTZ black hole with metric function given by $(7)$. A
massless scalar field perturbation in this space time satisfies the
Klein-Gordon equation,
\begin{equation}\label{}
     \frac{1}{\sqrt{-g}}\frac{\partial}{\partial x^{a}}\left((g^{ab}\sqrt{-g}\frac{\partial}{\partial
    x^{b}}\right)\Phi=0,\\
\end{equation}
which on expanding gives,
\begin{equation}\label{}
    \frac{1}{f(r)}\frac{\partial^{2}\Phi}{\partial t^2}-\frac{\partial}{\partial
    r}f(r)\frac{\partial \Phi}{\partial
    r}-\frac{1}{r^{2}}\frac{\partial^2 \Phi}{\partial \phi^2}=0.
\end{equation}
The metric function $f(r)$ is given by $(7)$. To separate the
angular variables, we make use of the ansatz,
\begin{equation}\label{}
    \Phi = \frac{R(r)}{r}e^{-i\omega t} e^{i m_l\phi},
\end{equation}
where $\omega$ is the frequency, $m_l$ is the angular momentum
quantum number. Using the above ansatz, the Klein-Gordon equation
can be re-written as,
\begin{equation}\label{}
    \frac{d^2 R}{dr^2} + \frac{f'(r)}{f(r)} \frac{dR}{dr} +
    \left[\frac{\omega^2}{f(r)^2}-\frac{(\frac{m_l^2}{r^2}-\frac{2 Q}{r^2}+\frac{c c_1
    m}{r})}{f(r)}\right] R = 0.
\end{equation}
Quasi normal modes are in going waves at the event horizon and
outgoing waves at the cosmological horizon, leading to the boundary
condition,
\begin{equation}
    R\rightarrow
    \begin{cases}
    e^{i\omega x}, &\text{as } x\rightarrow\infty\\
    e^{-i\omega x}, &\text{as } x\rightarrow-\infty
    \end{cases}
\end{equation}
Making a variable change $r\rightarrow 1/\xi$, the wave equation
becomes,
\begin{equation}\label{}
    \frac{d^2R}{d\xi^2} + \frac{p'}{p}\frac{dR}{d\xi} + \left[
\frac{\omega^2}{p^2} - \frac{2Q + \frac{2\Lambda}{\xi^2} - \frac{c
c_1 m}{\xi} - m_l^2}{p}\right]R=0,
\end{equation}
where,
\begin{eqnarray}
  p &=& M\xi^2 - c c_1 m\xi + 2Q \xi^2\ln{(\frac{1}{\alpha\xi})} + \Lambda, \\
  p' &=& 2(M-Q)\xi - c c_1 m + 4Q \xi\ln{(\frac{1}{\alpha\xi})}.
\end{eqnarray}
The wave equation given by $(13)$ has got the singularities at the
event horizon and at an outer horizon. In order to solve the wave
equation, the singularities have to be scaled out. Here, we first
scale out the divergent behavior at the outer horizon and then
re-scale to avoid the event horizon. To scale out the divergence at
outer horizon, we take\cite{35},
\begin{equation}\label{}
    R(\xi)=e^{i\omega \xi}u(\xi),
\end{equation}
where,
\begin{equation}
    e^{i\omega \xi}=(\xi-\xi_1)^{\frac{i\omega}{2\kappa_1}}%
    (\xi -\xi
    _2)^{\frac{i\omega}{2\kappa_2}},
\end{equation}
and,
\begin{equation}\label{}
    \kappa_i = \frac{1}{2}\frac{\partial f}{\partial r}\mid r\rightarrow
    r_i,
\end{equation}
is the surface gravity at each horizon. The master equation then
will take the form,\\ \\
\begin{equation}\label{}
    p u'' + (p' - 2 i\omega) u' - \left(2Q - \frac{2\Lambda}{\xi^2} - \frac{c c_1 m}{\xi} -
    m_l^2\right)u = 0
\end{equation}
This can be viewed  as,
\begin{equation}\label{}
    u''=\lambda_0(\xi)u' +s_0(\xi)u,
\end{equation}
with,
\begin{eqnarray}
  \lambda_0 &=& -\frac{(p' - 2 i\omega)}{p}, \\
  s_0 &=& \frac{\left(2Q - \frac{2\Lambda}{\xi^2} - \frac{c c_1 m}{\xi} -
    m_l^2\right)}{p}.
\end{eqnarray}
We employ the Improved Asymptotic Iteration Method (Improved AIM)
explained in Ref.\cite{36,37,38}. The coefficients are found out
upto $(n+2)^{th}$ derivative of $u$. It is assumed that when $n$ is
large the ratio of the derivatives,
$\frac{u^{n+2}}{u^{n+1}}=\frac{\lambda_{n-1}}{\lambda_n}$ converges
to a constant value, $\alpha$. This makes the quantization condition
given by,
\begin{equation}\label{}
    \lambda_n(x)s_{n-1}(x)-\lambda_{n-1}(x)s_n(x)=0,
\end{equation}
a possible one. It can be seen from $(21)$ that $\lambda_0$ contains
the quasi normal frequencies. So, the quantization condition given
by $(23)$ can be used to determine the quasinormal frequencies of
the black hole.
\begin{table*}
\caption{QNMs of linearly charged BTZ black hole for different
values of the massive parameter for dS space time with $Q=0.25$}
\centering
\begin{tabular*}{\textwidth}{@{\extracolsep{\fill}}|lr|lr|lr|@{}}
  \hline
   &$m=1$ &&$m=1.05$ &&$m=1.1$\\
  \hline
  $\Lambda$ & $\omega=\omega_R+\omega_I$ &$\Lambda$& $\omega=\omega_R+\omega_I$ &$\Lambda$ & $\omega=\omega_R+\omega_I$\\
  \hline\hline
  0.05  & 1.10826 - 0.11372 i& 0.13& 2.34196 - 0.20795 i& 0.19  & 4.17236 - 1.06710 i\\
  0.06  & 1.11679 - 0.12335 i& 0.15& 2.44520 - 0.20739 i& 0.21  & 4.27438 - 1.24924 i\\
  0.07  & 1.12571 - 0.13415 i& 0.17& 2.54513 - 0.20066 i& 0.23  & 4.36708 - 1.46493 i\\
  0.08  & 1.13484 - 0.14622 i& 0.19& 2.63890 - 0.19441 i& 0.25  & 4.44633 - 1.71966 i\\
  0.09  & 1.14390 - 0.15966 i& 0.21& 2.72810 - 0.19347 i& 0.27  & 4.50586 - 2.02037 i\\
  0.10  & 1.15262 - 0.17447 i& 0.21& 2.72810 - 0.19347 i& 0.28  & 4.52540 - 2.19059 i\\
  \hline
  0.11  & 1.57071 - 0.17563 i& 0.22& 2.77213 - 0.19538 i& 0.29  & 0.54837 - 2.25384 i\\
  0.12  & 1.57561 - 0.17204 i& 0.23& 2.81604 - 0.19882 i& 0.30  & 0.50212 - 2.32018 i\\
  0.13  & 1.58260 - 0.16856 i& 0.25& 2.90460 - 0.20967 i& 0.31  & 0.45065 - 2.39119 i\\
  0.14  & 1.59172 - 0.16559 i& 0.27& 2.99477 - 0.22423 i& 0.32  & 0.39409 - 2.46833 i\\
  0.15  & 1.60312 - 0.16277 i& 0.29& 3.08568 - 0.24039 i& 0.33  & 0.33391 - 2.55151 i\\
  0.16  & 1.61655 - 0.16277 i& 0.31& 3.25912 - 0.27670 i& 0.34  & 0.26958 - 2.63657 i\\
  \hline
\end{tabular*}
\end{table*}
\\ \\In Table $1$ we list the Quasi normal frequencies of the black hole
in the de Sitter space time for $m=1$, $m=1.05$ and $m=1.1$ for
different values of the cosmological constant. We have used the
parameter values $Q=0.25$, $m_l=1$, $\alpha=1$, $c=1$ and $c_1=1$.
In the numerical calculations we have used $15$ iterations. It is
observed that the behavior of the quasi normal frequencies change
after a particular $\Lambda$ value. This change in behavior is shown
in the table by a horizontal line as a separator. This sudden change
in behavior happens at $\Lambda=0.1$ for $m=1$, at $\Lambda=0.21$
for $m=1.05$ and at $\Lambda=0.28$ for $m=1.1$. The variation of the
QNMs with $\Lambda$ is shown in Fig. $1$.
\begin{figure}[h]
  \includegraphics[width=15pc]{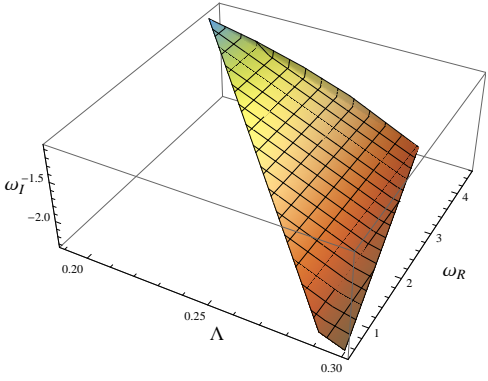}\\
  \caption{The behavior of QNMs with $\Lambda$ for $m=1.1$}
\end{figure}
The behavior of the QNMs for $m=1,0.05$ and $1.1$ given by Table $1$
are plotted in Fig. $2$.
\begin{figure}[h]
  \includegraphics[width=4.3cm]{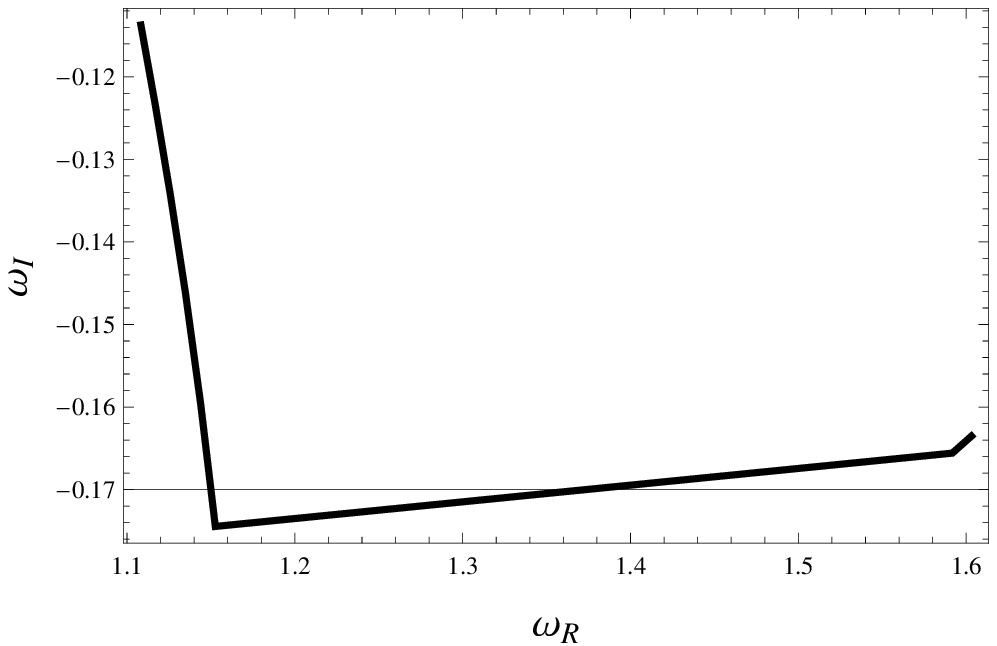}\includegraphics[width=4.3cm]{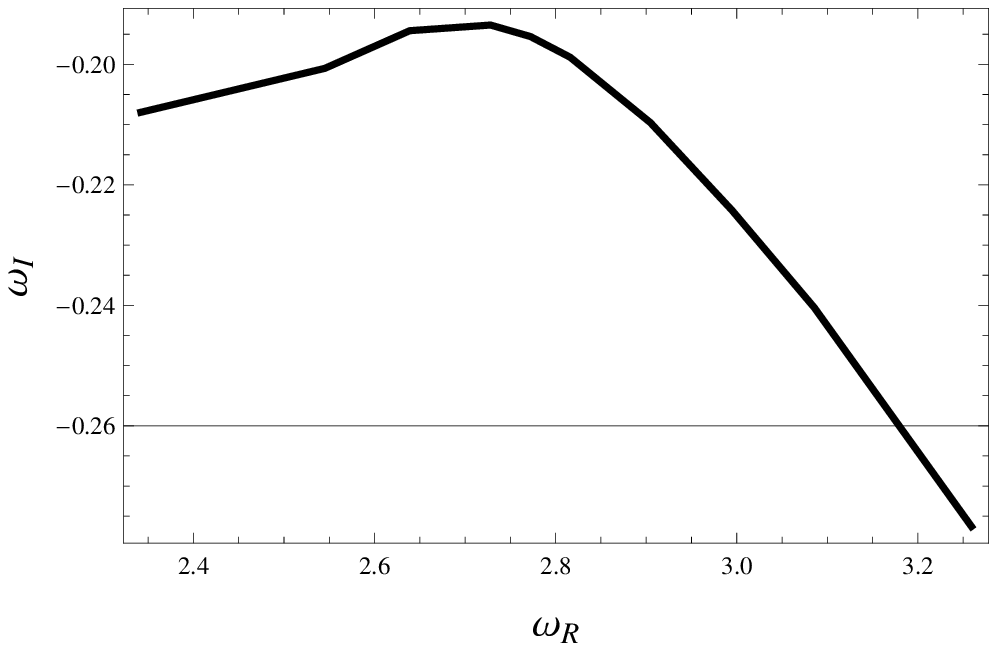}\\
  \begin{center}
  \includegraphics[width=4.3cm]{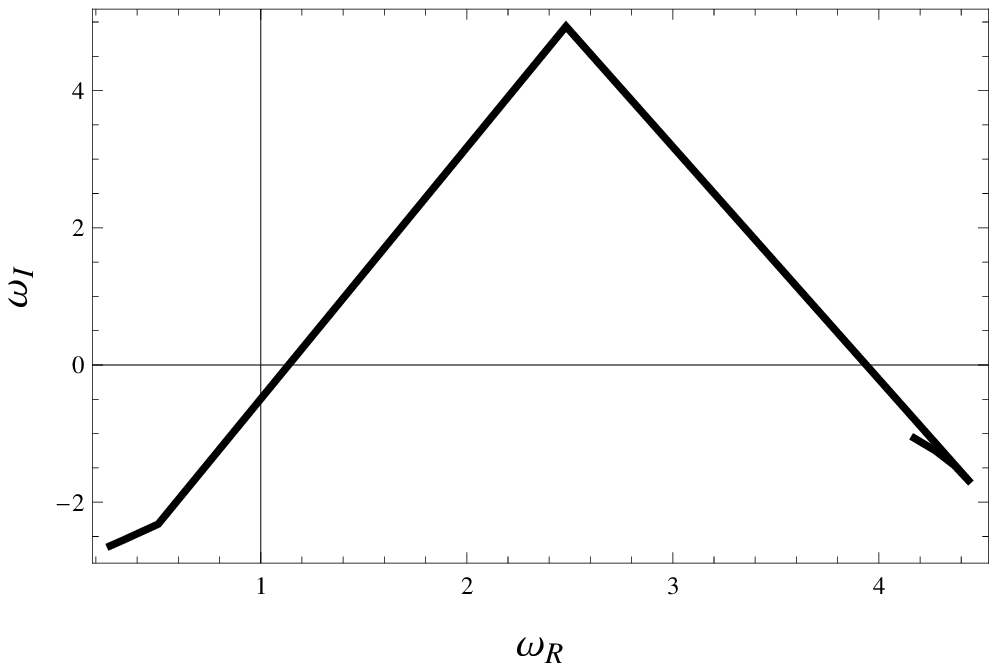}\\
  \caption{QNM behavior for linearly charged BTZ black hole for the massive parameter value $m=1,1.05,1.1$.
  The sudden change in the slope can be treated as an indicative of a possible phase transition.}
  \end{center}
\end{figure}
From the figures it can be clearly seen that the slope of the curve
changes suddenly at some transition point for $m=1,1.05,1.1$.  This
behavior can be treated as a clear indication of a phase transition.
However for the same values of the constant parameters this phase
transition occurs at different values of $\Lambda$ for the different
$m$ values. The higher the value of $m$, the larger
the value of $\Lambda$ at which the phase transition occurs.\\ \\
\begin{table*}[ht]
\caption{QNMs of linearly charged BTZ black holefor different values
of the massive parameter for dS space time with $Q=0.35$} \centering
\begin{tabular*}{\textwidth}{@{\extracolsep{\fill}}|lr|lr|lr|@{}}
  \hline
   &$m=0.9$ &&$m=0.95$ &&$m=1.0$\\
  \hline
  $\Lambda$ & $\omega=\omega_R+\omega_I$ &$\Lambda$& $\omega=\omega_R+\omega_I$ &$\Lambda$ & $\omega=\omega_R+\omega_I$\\
  \hline\hline
  0.09  & .898571 - .0783066i   & 0.05  &1.16072 - .0686295i    &0.01   & 1.68971 - .172263i\\
  0.10  & .901026 - .0798863i   & 0.06  &1.18398 - .0690504i    &0.015  & 1.69779 - .581786i\\
  0.11  & .902855 - .0851571i   & 0.07  &1.20614 - .0630027i    &0.02   & 1.70293 - .198065i\\
  0.12  & .902565 - .0947457i   & 0.08  &1.21651 - .0514301i    &0.025  & 1.70506 - .214255i\\
  0.13   & 1.03909 - .107478i   & 0.09  &1.21431 - .0415960i    &0.03   & 1.70405 - .232616i\\
  \hline
  0.14  & 1.01865 - .0859070i   & 0.10  &1.20180 - .0347956i    &0.04   & 1.69178 - .27573i\\
  0.15  & 1.00159 - .0683552i   & 0.11  &1.17941 - .0303883i    &0.05   & 1.66379 - .327079i\\
  0.16  & .983281 - .0666629i   & 0.12  &1.14639 - .0274678i    &0.06   & 1.61649 - .386106i\\
  0.17  & .961166 - .0464981i   & 0.13  &1.10110 - .0251206i    &0.07   & 1.54442 - .451769i\\
  0.18  & .933873 - .0396936i   & 0.14  &1.04093 - .0224859i    &0.08   & 1.43902 - .521872i\\
  \hline
\end{tabular*}
\end{table*}
In Table $2$, we have shown the Quasi normal frequencies for
$Q=0.35$ for $m=0.9$, $m=0.95$ and $m=1.0$ with the parameter values
$m_l=c=c_1=1$. The behavior of these QNMs are shown in Figure $3$.
Just like in the case where $Q=0.25$, here also there is a sudden
change in the slope of the curve after a particular $\Lambda$
indicating that of a phase transition.\\
\begin{figure}[h]
  \includegraphics[width=4.3cm]{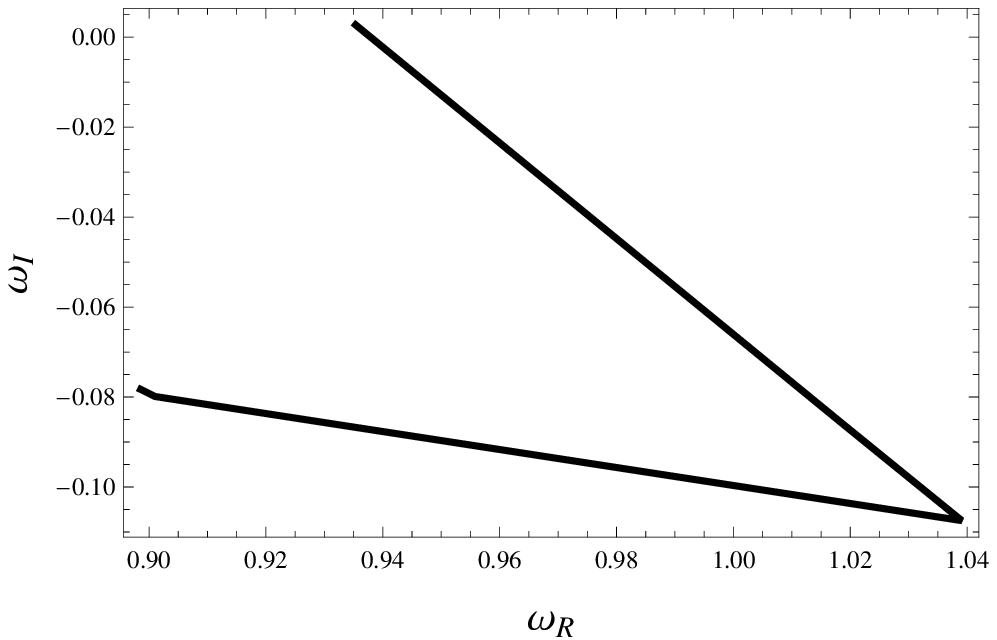}\includegraphics[width=4.3cm]{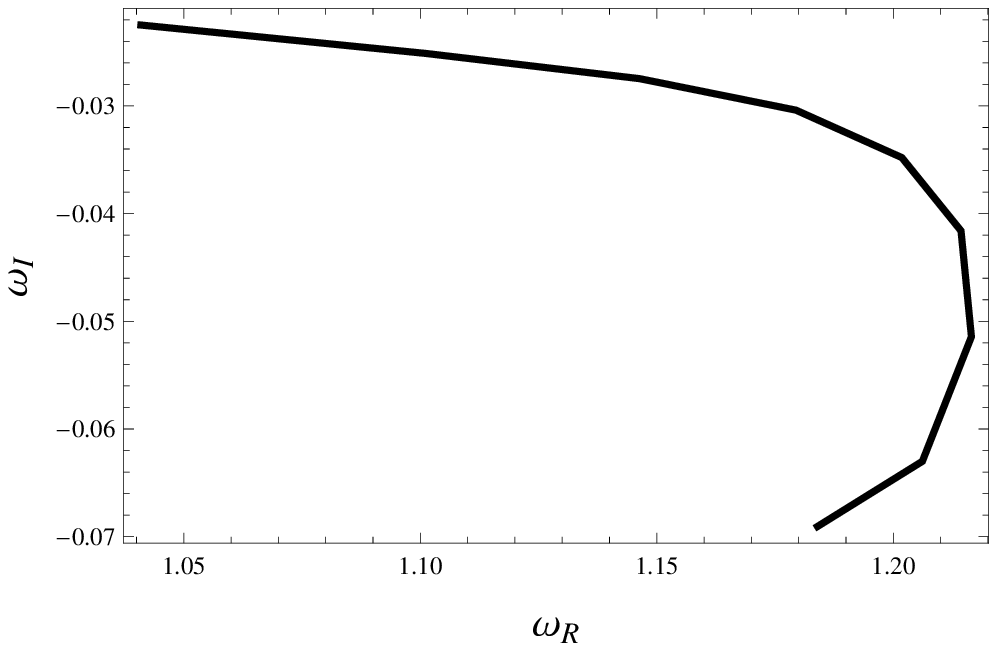}\\
  \begin{center}
  \includegraphics[width=4.3cm]{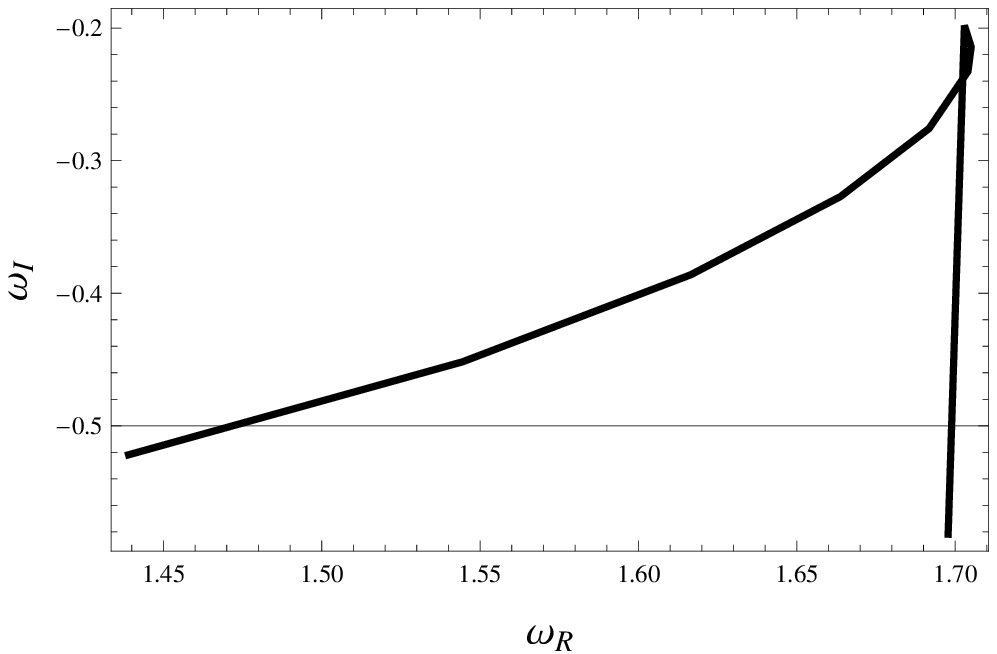}
  \caption{QNM behavior for linearly charged BTZ black hole for dS spece time with $Q=0.35$ for the massive parameter value $m=0.9,0.95,1.0$.
  The sudden change in the slope can be treated as an indicative of a possible phase transition.}
  \end{center}
\end{figure}
Thus for both values of $Q$ the black hole shows phase transition.
We  can  see that for the value $m=1.0$ the phase transition happens
at a different value of $\Lambda$ compared to $Q=0.25$ and $Q=0.35$
cases.
\begin{figure}[h]
  \includegraphics[width=15pc]{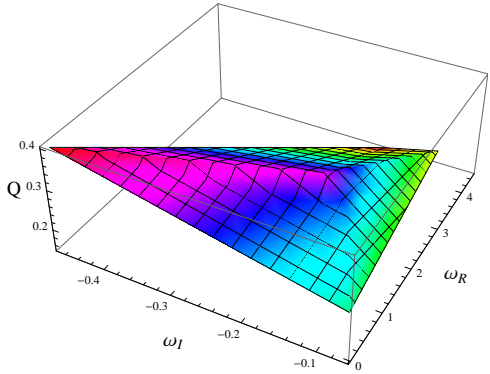}\\
  \caption{Variation of QNMs with charge $Q$ for the dS space time}
\end{figure}
In Table $3$ we show the QNMs for an AdS space time for the
parameter values $Q=0.1$, $\alpha=1$, $c=1$, $c_1=1$ and
$m=1,1.05,1.1$. From the table, it can be observed that the
$\omega_R$ and $\omega_I$ continuously decrease and after reaching a
particular point $(\Lambda=0.12)$, the real part suddenly increases
and then continuously decrease whereas the imaginary part continues
to decrease. This jump can be treated as an indication of an
inflection point.
\begin{table*}[ht]
\centering \caption{QNMs of linearly charged BTZ black hole for
different values of the massive parameter  for AdS space time with
$Q=0.1$ }
\begin{tabular*}{\textwidth}{@{\extracolsep{\fill}}|lr|lr|lr|@{}}
  \hline
   &$m=1.0$ &&$m=1.05$ &&$m=1.1$\\
  \hline
  $\Lambda$ & $\omega=\omega_R+\omega_I$ &$\Lambda$& $\omega=\omega_R+\omega_I$ &$\Lambda$ & $\omega=\omega_R+\omega_I$\\
  \hline\hline
  -0.06 & 1.83077 - 5.78701 i &-0.05  &1.39873 - 7.68495 i    & &\\
  -0.07 & 1.70014 - 5.33444 i &-0.06  &1.29210 - 7.27705 i    &-0.04 &0.75408 - 9.41718 i  \\
  -0.08 & 1.53828 - 4.92198 i &-0.07  &1.14457 - 6.93423 i    &-0.05 &0.63741 - 9.08170 i \\
  -0.09 & 1.34563 - 4.54476 i &-0.08  &0.95604 - 6.62556 i    &-0.06 &0.48892 - 8.73203 i \\
  -0.10 & 1.11762 - 4.20206 i &-0.09  &0.72592 - 6.35819 i    &-0.07 &0.21318 - 8.39613 i\\
  -0.11 & 0.84041 - 3.89983 i &-0.95  &0.58718 - 6.23146 i    &&\\
  -0.12 & 0.48197 - 3.66706 i &-0.10  &0.40624 - 6.12793 i &&\\
  \hline
  -0.13 & 0.81813 - 4.06506 i &-0.11  &1.57334 - 7.10865 i    &-0.08 &2.18254 - 10.2207 i\\
  -0.135& 0.75562 - 3.41486 i &-0.13  &1.12639 - 5.99601 i    &-0.09 &1.44272 - 10.1043 i\\
  -0.14 & 0.32251 - 2.91165 i &-0.14  &0.86214 - 5.07753 i    &-0.10 &1.41871 - 9.40952 i \\
  \hline
\end{tabular*}
\end{table*}
The $\omega_R$ versus $\omega_I$ is plotted in Fig. $3$. It can be
seen from the figure that there is no drastic change in the slope
and the behavior of the QNMs are same. Hence it can be inferred that
there will be no phase transition.
\begin{figure}[h]
  \includegraphics[width=4.3cm]{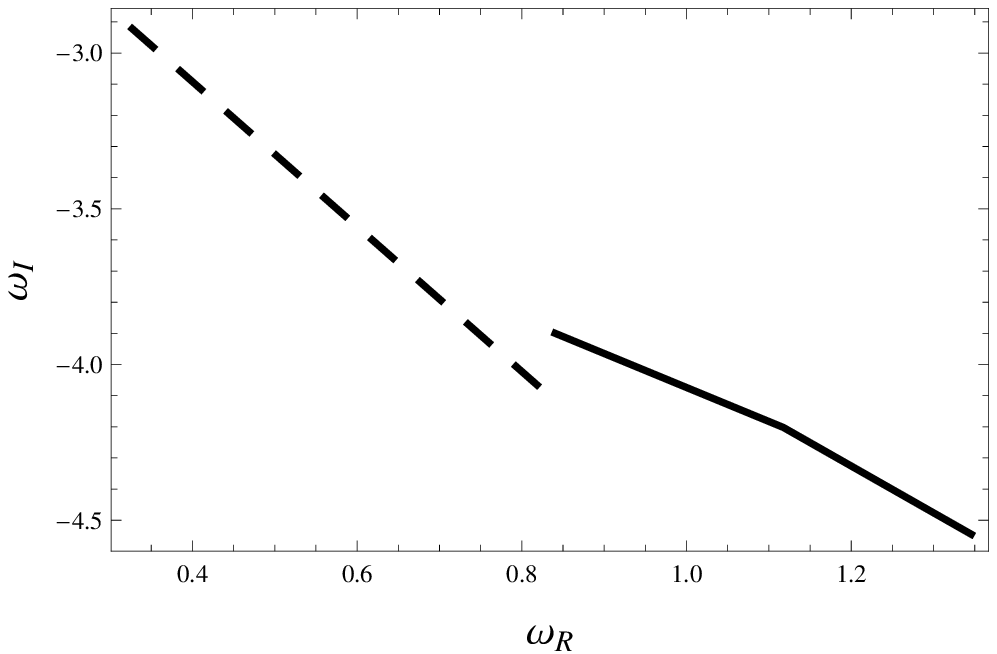}\includegraphics[width=4.3cm]{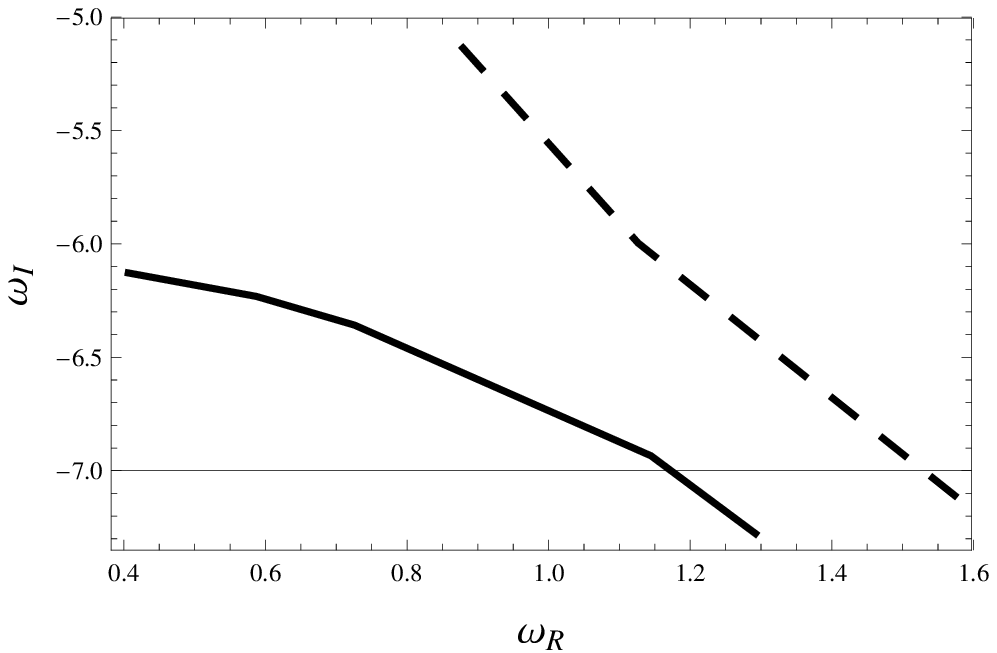}\\
  \begin{center}
  \includegraphics[width=4.3cm]{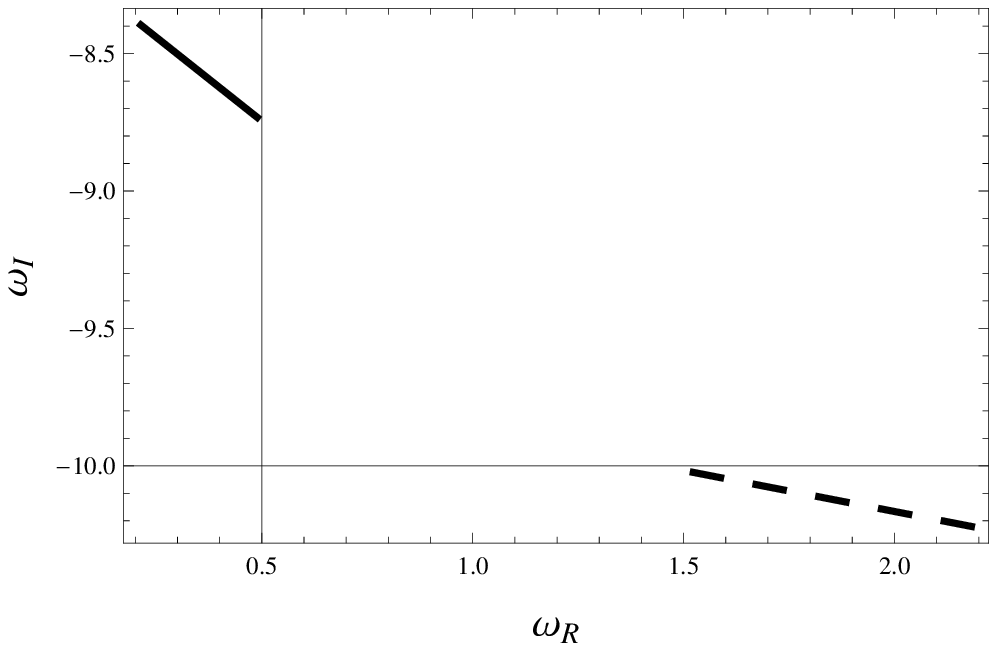}\\
  \caption{QNM behavior for linearly charged BTZ black hole with  charge Q =0.1 for AdS space time for the massive parameter value $m=1,1.05,1.1$. The dotted line represents  the
  behavior of QNMs after the inflection point.
  The behavior of QNMs are seen to be the similar in the plots. There is no much difference in the slope of the curves}
  \end{center}
\end{figure}
In Table $4$ we have calculated the QNMs for the AdS space time for
the charge $Q=0.25$. Fig. $6$ shows the behavior of quasi normal
frequencies for the above case. It can be seen that there is a
sudden change in slope of the curve after reaching a particular
$\Lambda$ indicating a phase transition. For $Q=0.1$ the AdS black
hole space time did not show any phase transition behavior but for
$Q=0.25$ it is found to be showing a phase transition behavior.
Hence, it can be inferred that the phase transition behavior depends
on the charge $Q$.
\begin{table*}
\caption{QNMs of linearly charged BTZ black hole for different
values of the massive parameter for AdS space time with $Q=0.25$}
\centering
\begin{tabular*}{\textwidth}{@{\extracolsep{\fill}}|lr|lr|lr|@{}}
  \hline
   &$m=0.95$ &&$m=1.0$ &&$m=1.05$\\
  \hline
  $\Lambda$ & $\omega=\omega_R+\omega_I$ &$\Lambda$& $\omega=\omega_R+\omega_I$ &$\Lambda$ & $\omega=\omega_R+\omega_I$\\
  \hline\hline
  0.01  & .292587 - 9.19482i    & 0.13  & .820054 - 4.96149i    &0.29   & 1.01098 - .0351877i\\
  0.02  & .772162 - 8.39220i    & 0.15  & .431983 - 3.38060i    &0.31   & 1.02868 - .0148701i\\
  0.03  & .844245 - 7.75702i    & 0.17  & .00879691 - .0348464i &0.32   & 1.04119 - .00215093i\\
  \hline
  0.04  & .820904 - 7.24016i    &0.19   & 1.72429 - .0660172i   &0.33   &1.62431 - .102106i\\
  0.05  & .759655 - 6.75177i    &0.20   & 1.73419 - .0561830i   &0.34   &1.62400 - .083530i\\
  0.07  & .551068 - 5.89321i    &0.21   & 1.74461 - .043151i    &0.35   &1.61905 - .067770i\\
  0.09  & .390010 - 4.86350i    &0.22   & 1.75581 - .0330092i   &0.36   &1.60957 - .060276i\\
  0.11  & .243717 - 3.77402i    &0.23   & 1.76769 - .0186321i   &&\\
  \hline
\end{tabular*}
\end{table*}
\begin{figure}[h]
  \includegraphics[width=4.3cm]{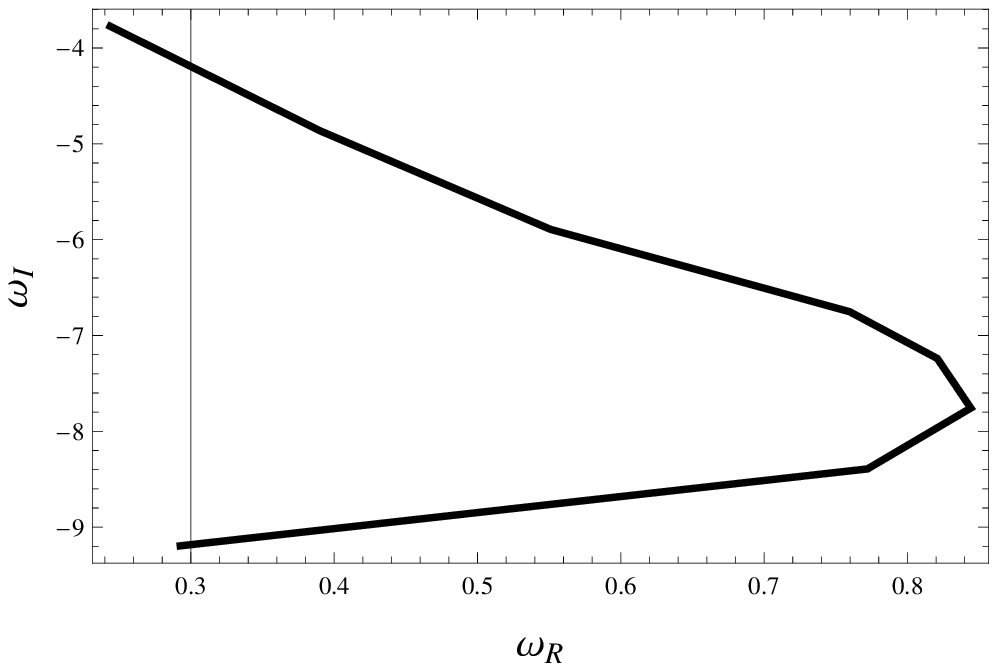}\includegraphics[width=4.3cm]{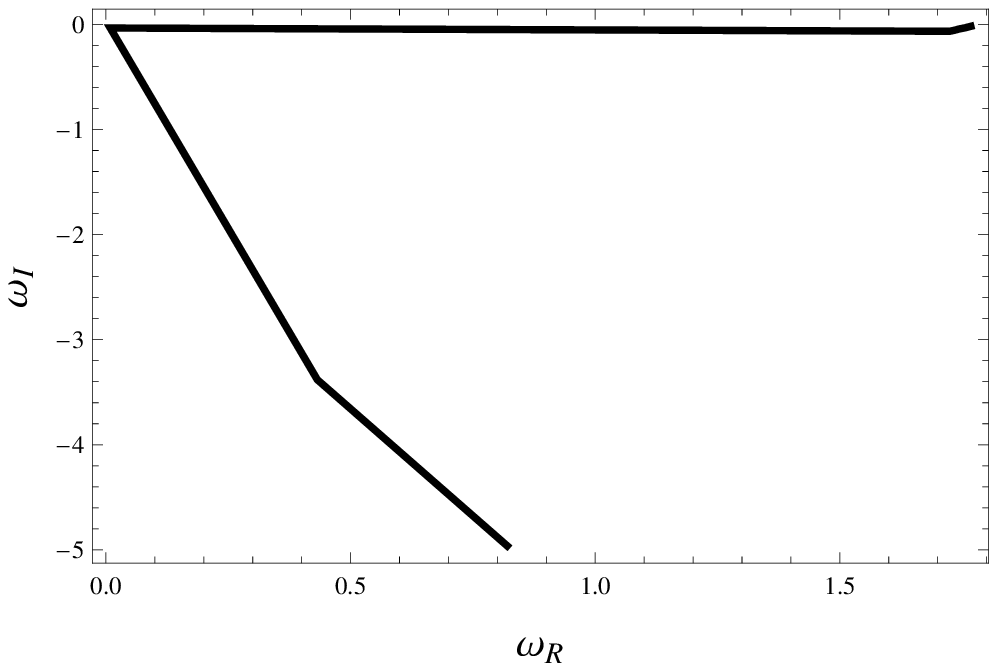}\\
  \begin{center}
  \includegraphics[width=4.3cm]{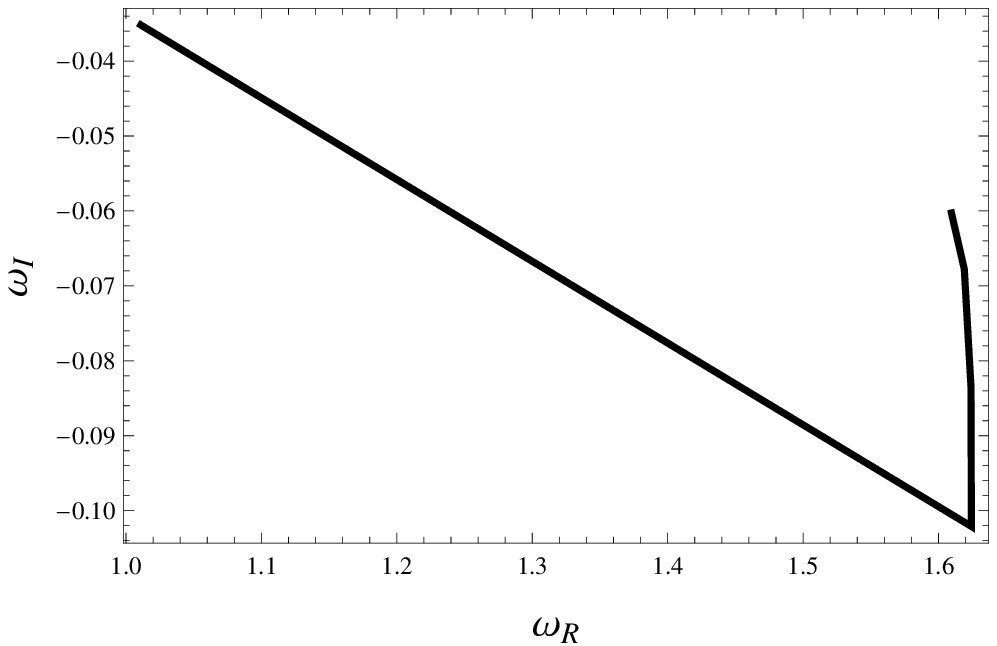}
  \caption{QNM behavior for linearly charged BTZ black hole for $Q=0.25$ the massive parameter value $m=0.95,1.0,1.05$.}
  \end{center}
\end{figure}
\\ \\Now, it would be interesting to check the variation of QNMs with $Q$.
Table $5$ shows the variation of quasi normal frequencies with
charge $Q$ for dS space time. It can be seen that the behavior of
quasi normal frequency changes frequently. The phase transition
behavior is highly dependent on the charge. The phase does not
remain the same for a wide range of charge and hence phase
transition is found to happen frequently over a range of charges.
\begin{table}
  \centering
  \caption{Table showing the variation of QNMs with $Q$ for dS space time}\label{}
  \begin{tabular}{|l|c|}
    \hline\hline
    Q & $\omega$ \\
    \hline
    0.15 & 4.47348 - 0.19884 i \\
    0.20 & 4.33915 - 0.108836 i \\
    \hline
    0.25 & 0.0930679 - 0.0668980 i \\
    0.30 & 1.54638 - 0.132502 i \\
    0.35 & 1.68971 - 0.172263 i \\
    \hline
    0.40 & 0.0325096 - 0.466834 i \\
    \hline\hline
  \end{tabular}
\end{table}
The variation of QNMs with charge for the AdS case is shown in Table
$6$.
\begin{table}
  \centering
  \caption{Table showing the variation of QNMs with $Q$ for the AdS space}\label{}
  \begin{tabular}{|l|c|}
    \hline\hline
    Q & $\omega$ \\
    \hline
    0.05    &1.12930 - 9.66585 i\\
    0.10    &1.14048 - 9.54589 i\\
    0.15    &2.42998 - 12.8629 i\\
    0.20    &3.49176 - 14.0316 i\\
    0.25    &3.64711 - 13.9835 i\\
    \hline
    0.30    &0.294699 - 9.56518 i\\
    0.35    &0.557735 - 9.42807 i\\
    0.40    &0.792729 - 9.21181 i\\
    0.45    &0.940783 - 9.04476 i\\
    0.50    &1.03930 - 8.88485 i\\
    \hline\hline
  \end{tabular}
\end{table}
It can be seen that compared to the dS case, phase transition does
not happen frequently, ie., the phases remain the same for most of
the values of charge and a transition happens only for certain small
range of charge values. This behavior can  be seen in Fig $7$.\\
\begin{figure}[h]
\includegraphics[width=15pc]{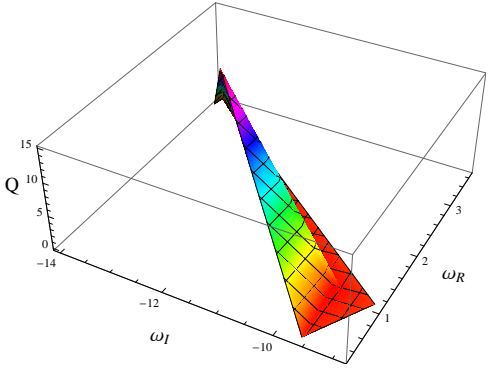}
  \caption{Variation of QNMs with charge $Q$ for AdS space time}
\end{figure}
\section{Thermodynamics of the black hole}
In this section, we study the thermodynamics of the linearly charged
BTZ black hole in the (Anti)de Sitter space time in massive gravity.
The mass of the black hole, $m_0$, is given by the solution of the
condition $f(r)|_{r\rightarrow r_H}=0$ as,
\begin{equation}\label{}
    m_0= m^2 c c_1 r_H+\Lambda r_{H}^2-2 Q \ln{(\frac{r_H}{\alpha})}.
\end{equation}
The temperature of the black hole is given by
$\frac{1}{4\pi}f'(r)|_{r\rightarrow r_{H}}$ which gives,
\begin{equation}\label{}
    T=\frac{c c_1 m^2}{4 \pi }-\frac{Q}{2 \pi  r_H}+4 P r_H.
\end{equation}
where $P=\frac{\Lambda}{8\pi}$. Finally, the entropy is evaluated
from the expression $S=\int_{0}^{r_H}\frac{1}{T}\frac{\partial
m_0}{\partial r}dr$ which gives,
\begin{equation}\label{}
    S=4\pi r_H.\\
\end{equation}
Then the equation of state, $P(V,T)$ can be obtained from the
expression for temperature, $(25)$, as,
\begin{equation}\label{}
    P=\frac{Q}{8 \pi  r_H^2}+\frac{-c c_1 m^2+4 \pi  T}{16 \pi  r_H}.
\end{equation}
For an $(n+2)$ dimensional massive gravity, the volume is given
by\cite{39}, $V=(\frac{\partial H}{\partial
P})_{S,Q}=\frac{V_n}{n+1}r^{n+1}$. With, $n=1$, the calculation
gives the horizon radius in terms of its volume as,
$r_H=(\frac{V}{8\pi})^{1/2}$.\\ \\
To specify the phase transition it will be useful to introduce the
Gibbs free energy as a Legendre transformation of enthalpy as,
\begin{equation}
    G=H-TS,
\end{equation}
where $H$ is the  enthalpy, $T$ is the temperature given by $(25)$
and $S$ is the entropy given by $(26)$. We use the black hole mass
$m_0$ as the enthalpy since $H\equiv m_0$ rather than the internal
energy of the gravitational system\cite{21}. Substituting $(24)$,
$(25)$ and $(26)$ in $(28)$, we get an expression for the Gibbs free
energy as,
\begin{equation}\label{12 a}
    G(T,\Lambda)=2Q + \Lambda r_H^2 - 2 Q \ln{(\frac{r_H}{\alpha})}.\\
\end{equation}

Fig. $8$ shows the  variation of Gibbs free energy  with temperature
plotted using $(25)$ and $(29)$. Top of the figure shows the G-T
plot for $P=\frac{\Lambda}{8\pi}=0.001$. It can be seen that the
upper branch which lies in the positive Gibbs free energy region
moves towards the lower branch which lies in the \lq positive
temperature-negative Gibbs free energy\rq\, region which indicates a
possible phase transition. The bottom plot shows variation of G with
T for $P=-0.001$. The plot lies in the positive Gibbs free energy
region and shows a cusp like behavior.
\begin{figure}[h]
\centering
  \includegraphics[width=16pc]{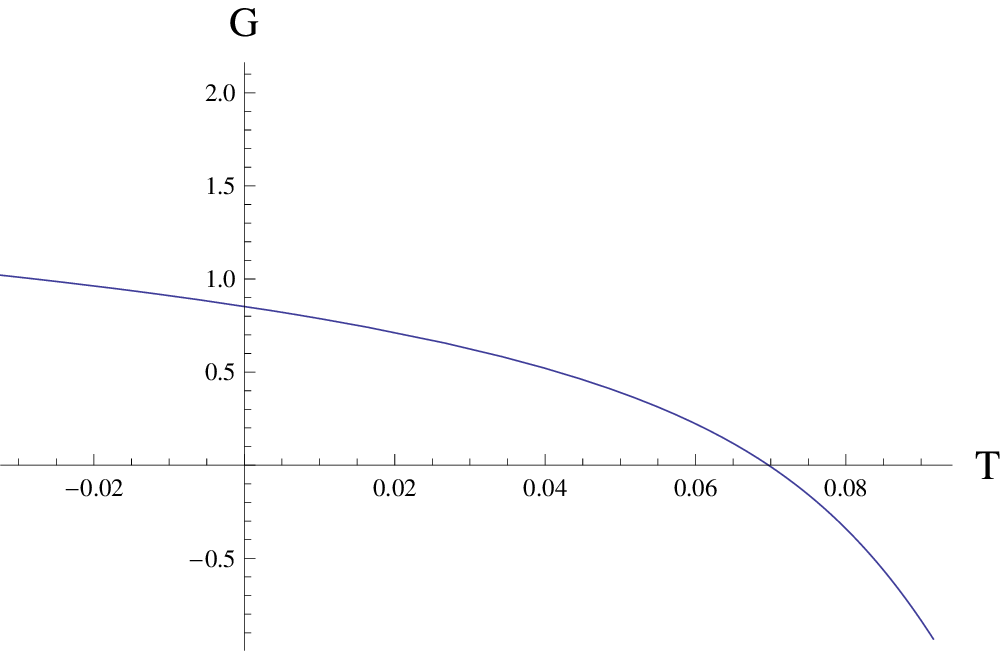}\\
  \includegraphics[width=16pc]{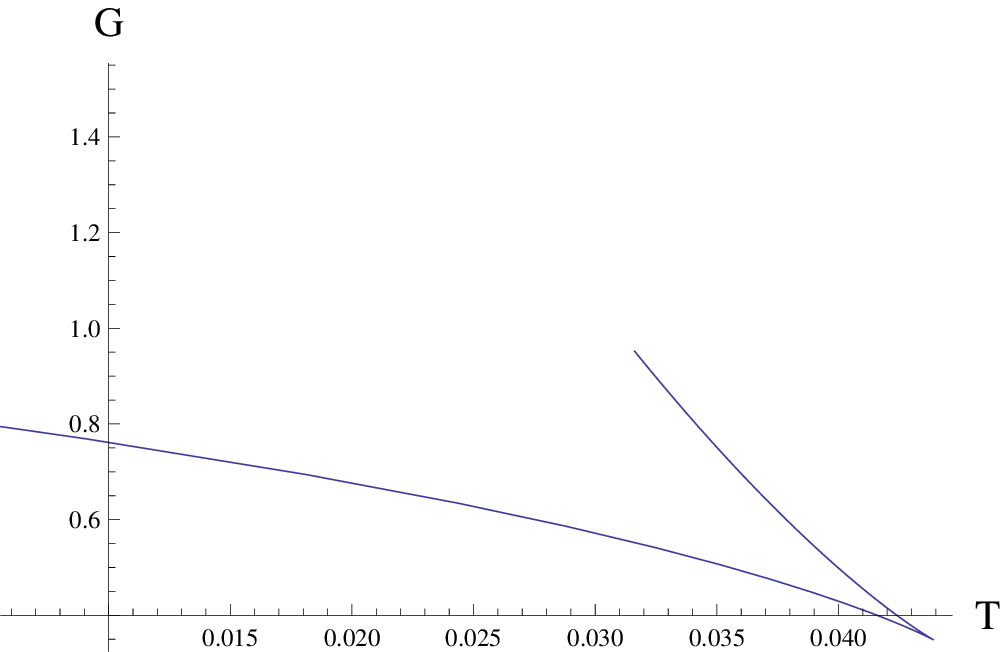}
  \caption{Variation of Gibbs free energy with temperature for
  the dS space time (top) and AdS space tome (bottom).}
\end{figure}

Fig. $9$ shows the variation of pressure, $P$ and temperature, $T$
with the horizon radius, $r_H$, for fixed values of temperature and
pressure respectively. Top of the Fig. $9$, shows the variation of
temperature with $r_h$ given by $(25)$ for the pressure values
$P=-0.003$, $-0.002,$ $-0.001,$ $0.001,$ and $0.002$. The bottom of
the Fig. $9$ shows the variation of pressure with $r_h$ given by
$(27)$ for the fixed values of temperature, $T=-0.3,$ $-0.2,$
$-0.1,$ $0.1,$ and $0.2$.

\begin{figure}[h]
  \includegraphics[width=15pc]{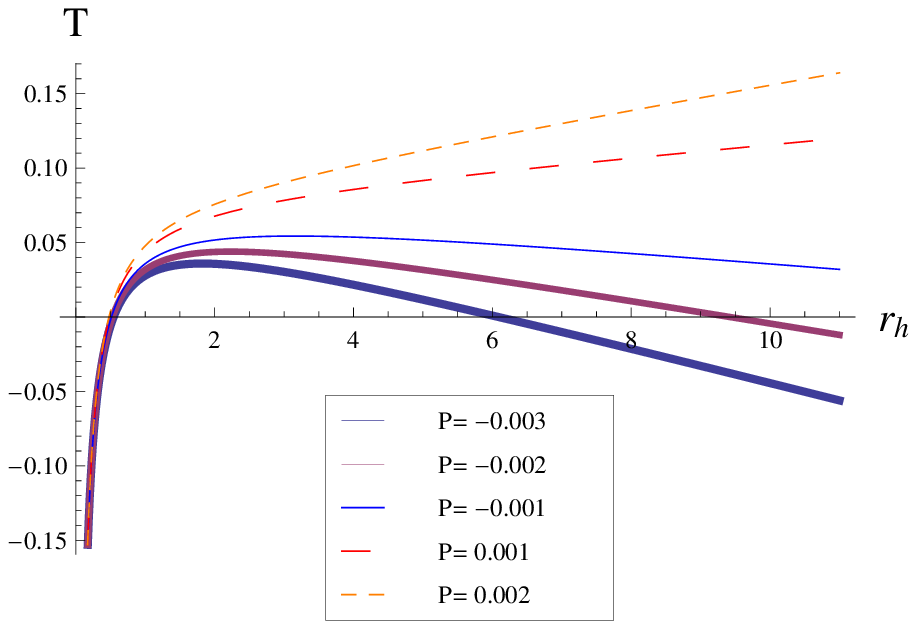}\hspace{0.5cm}\includegraphics[width=15pc]{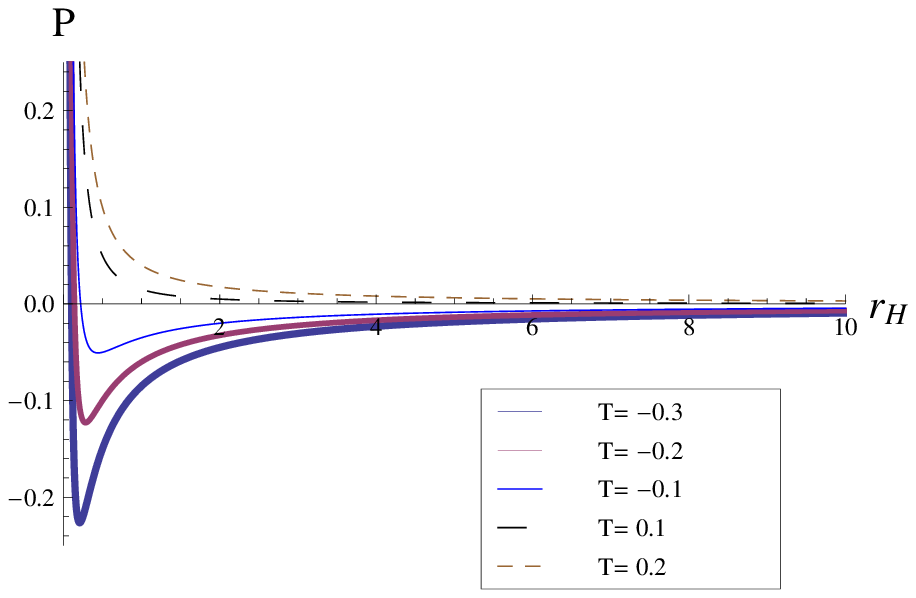}\\
  \caption{\textbf{Top : }Variation of $T$ with $r_H$. \textbf{Bottom : }Variation of $P$ with $r_H.$}
\end{figure}
More details regarding the phase transition can be extracted form
the entropy of the system. The temperature-entropy relation would be
worth looking at. For that the expression for $r_H$ derived from
$(26)$ is substituted into $(25)$ so that we get an expression
relating the entropy and temperature as,
\begin{equation}\label{}
    T=-\frac{2 Q}{S}+\frac{2\pi c c_1 m^2  + \Lambda\, S }{8 \pi
    ^2}.
\end{equation}
Fig. $10$ shows the $S-T$ plots for the values $\Lambda=0.1$ and
$\Lambda=-0.1$, with the parameter values $m_0=c=c_1=1$, $\alpha=1$,
$Q=0.25$ and $m=1$. It can be seen that $S$ remains positive only
for a small range of temperature and both of them show phase
transition behavior.
\begin{figure}[h]
\begin{minipage}{\columnwidth}
\includegraphics[width=17pc]{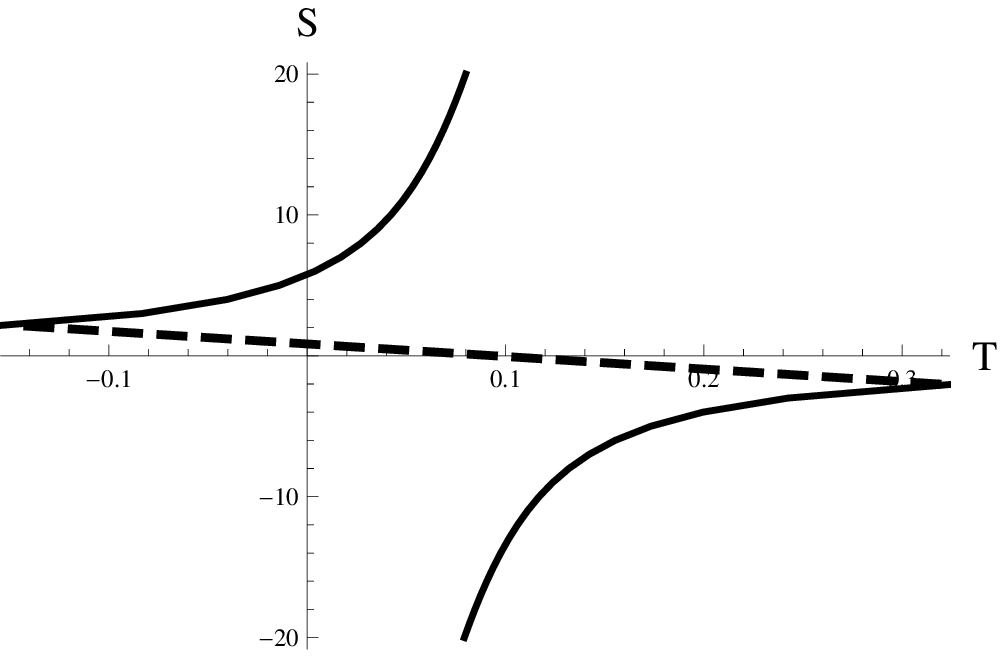}\\ \includegraphics[width=17pc]{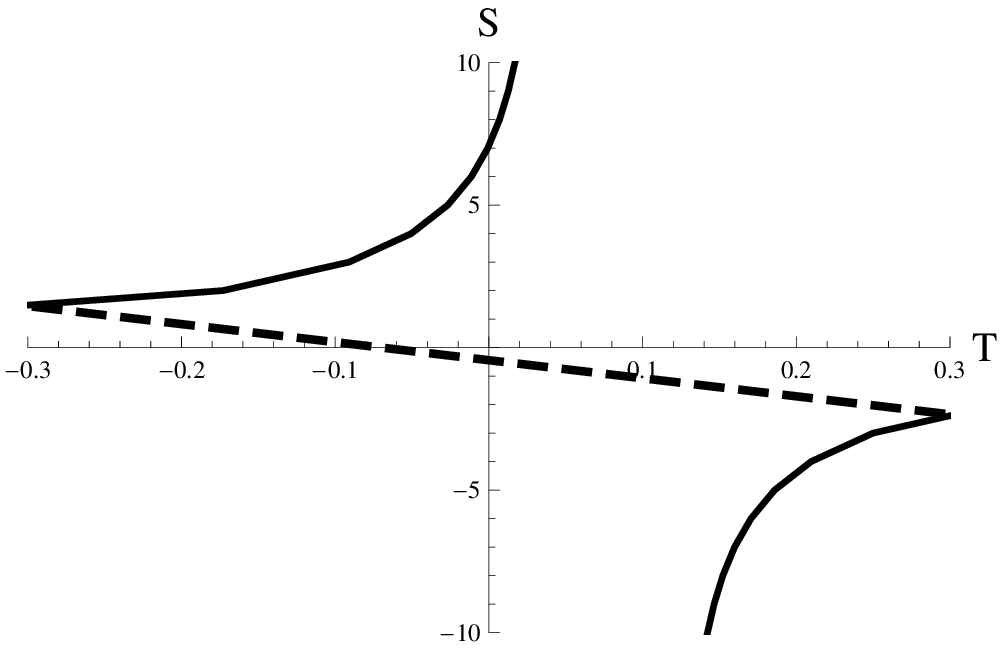}
\end{minipage}
\caption{The variation of thermal entropy with temperature for the
$\Lambda=0.1$(top) and $-0.1$(bottom) in dS space time}
\end{figure}
Now, in order to study the stability of the phases or the
feasibility of the above phase transitions, it may be worth looking
at the behavior of specific heat  with  temperature. If the behavior
of heat capacity indicates that as  the  temperature varies the heat
capacity  makes a  transition   from  negative values to  positive
values the system undergoes  a  phase  transition. Negative heat
capacity represents unstable state while  positive value  of
specific  heat implies  a stable state. The specific heat is given
by,
\begin{equation}\label{}
    C_Q=\frac{T}{(\frac{\partial T}{\partial S})_Q},
\end{equation}
which from $(26)$ and $(30)$ leads to,
\begin{equation}\label{}
    C_Q=2\pi r_h \frac{\left(-2 Q+r_h \left(m^2+2 r_h \Lambda \right)\right)}{Q+r_h^2 \Lambda }.
\end{equation}
The plots of specific heat versus temperature for $\Lambda=0.1$ and
$\Lambda=-0.1$ is given in Fig. $11$ for the parameter values
$m=c=c_1=1$ and $Q=0.25$ . From the plot it can be clearly
understood that for  $\Lambda=0.1$, the specific heat changes from
negative to positive values indicating a phase transition from
unstable to stable configuration. For  $\Lambda=-0.1$, from the
figure we can say that it somewhat shows a phase transition behavior
however, it is observed that for given constant parameter values,
the black holes in AdS space time show this phase transition
behavior only for a very small range of $\Lambda$ values whereas in
dS space time it shows phase transition for a wide range of
$\Lambda$ values.
\begin{figure}[h]
  \includegraphics[width=17pc]{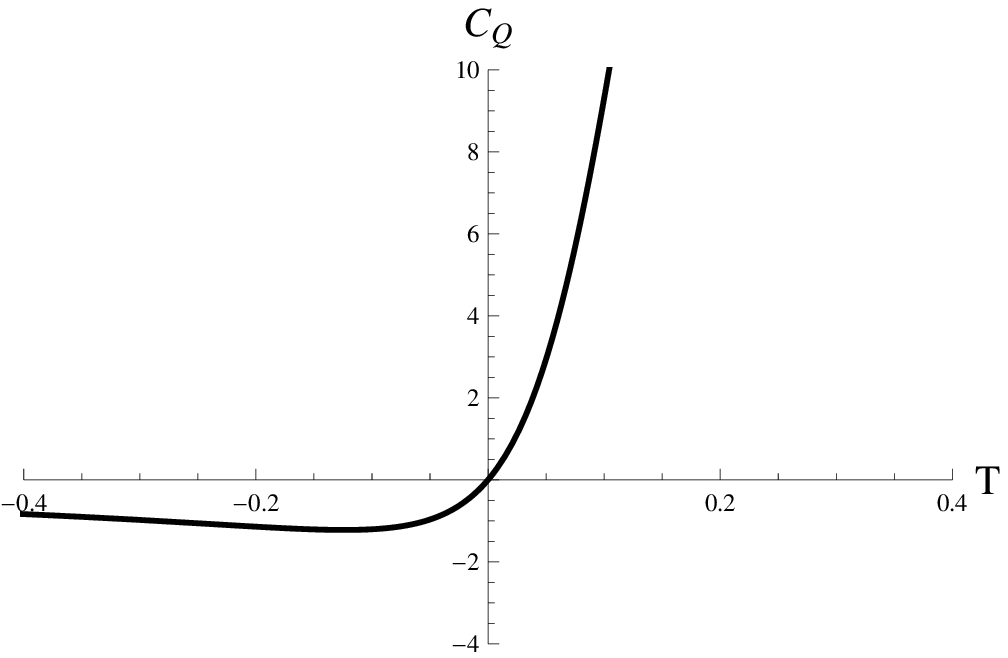}\\ \includegraphics[width=17pc]{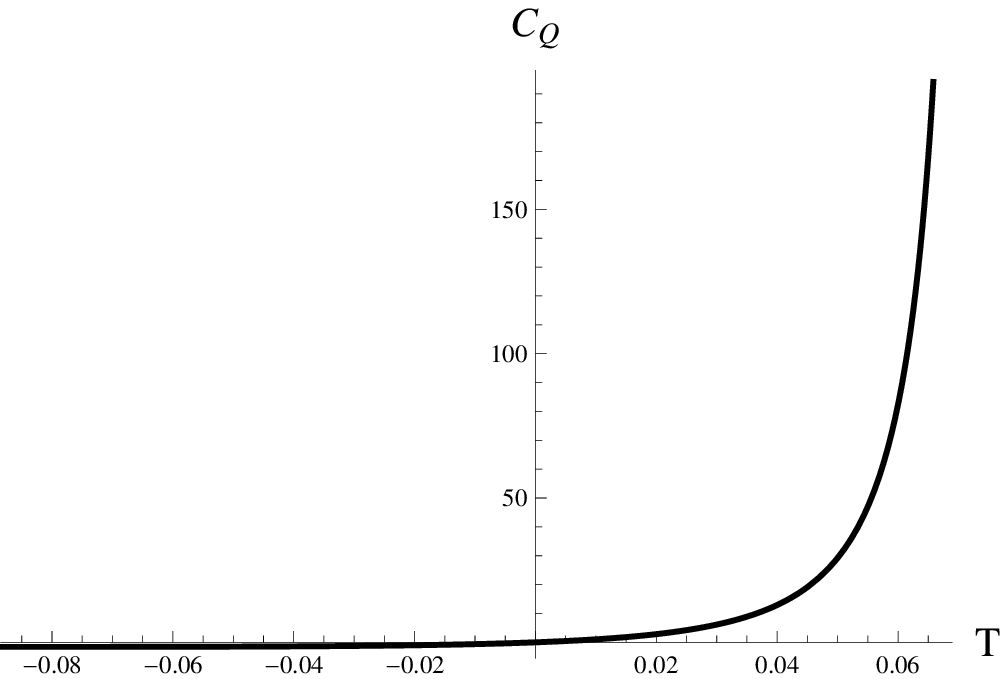}
  \caption{Figure showing the variation of specific heat with temperature for $\Lambda=0.1$(top) and
  $\Lambda=-0.1$(bottom). }
\end{figure}
It would also be worth noting that the variation of the behavior of
specific heat with $Q$. For this, we have plotted variation of
specific heat with temperature for $Q=0.1,0.25,0.5,0.6$ for dS space
time; the other parameters remaining the same and is shown in Fig
$12$.
\begin{figure}[h]
    \begin{minipage}{\columnwidth}
  \includegraphics[width=10pc]{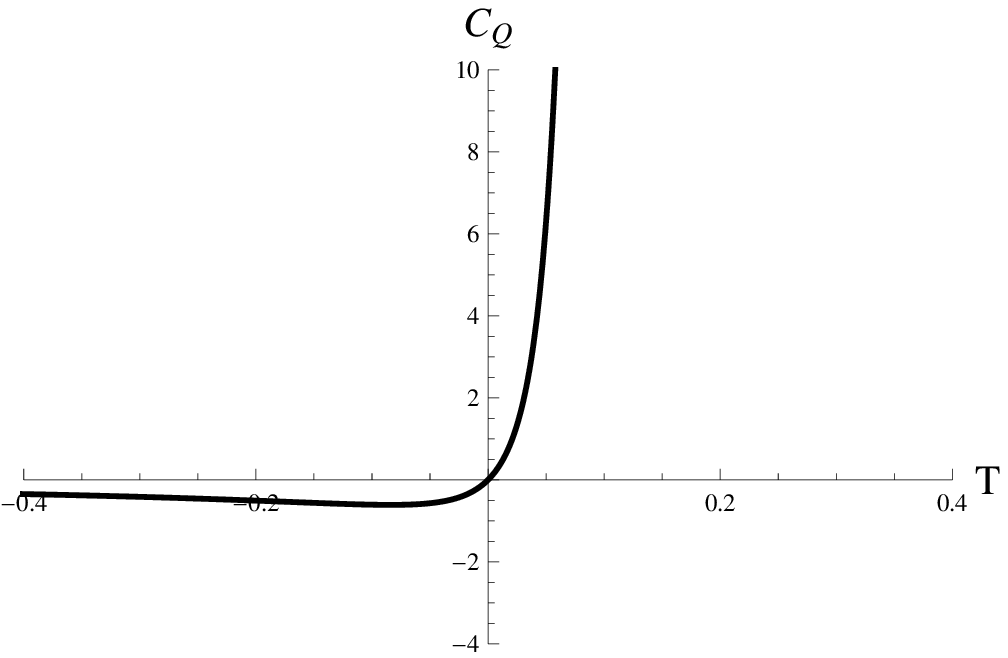}\hspace{0.5cm}\includegraphics[width=10pc]{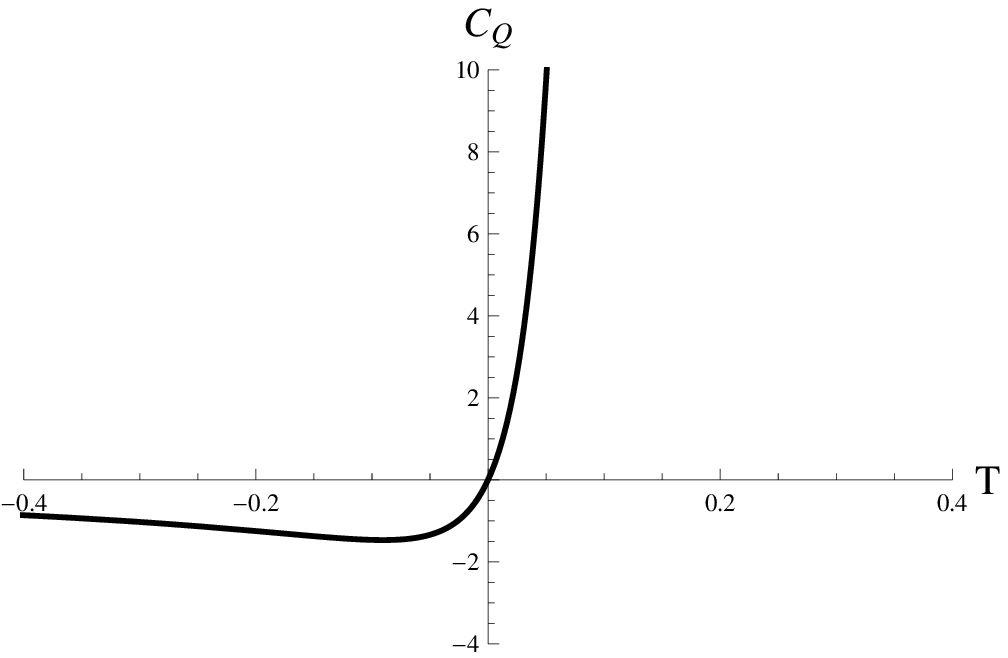}\\
  \includegraphics[width=10pc]{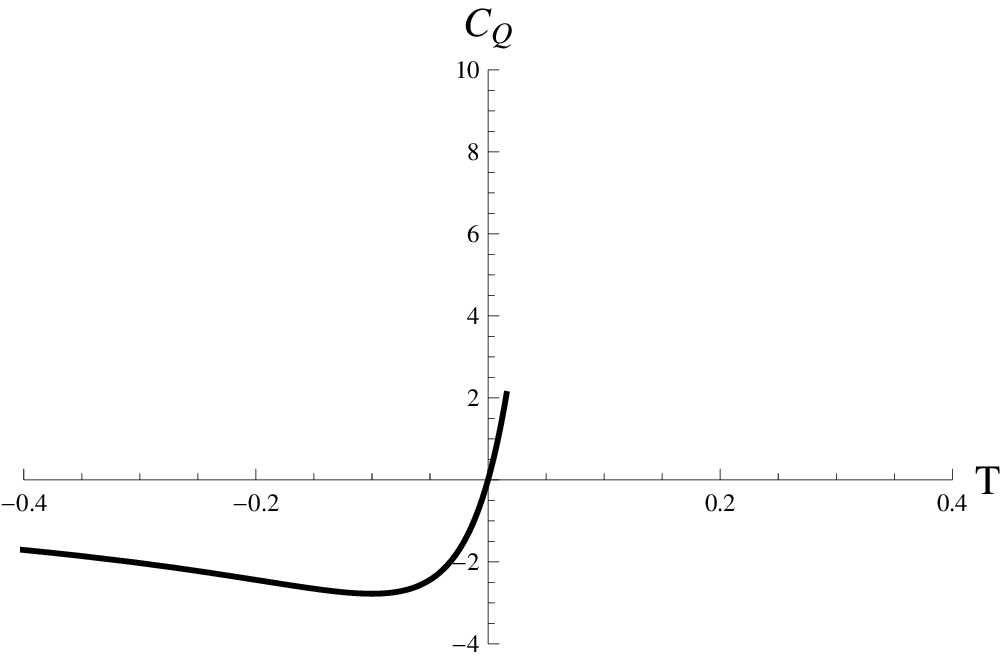}\hspace{0.5cm}\includegraphics[width=10pc]{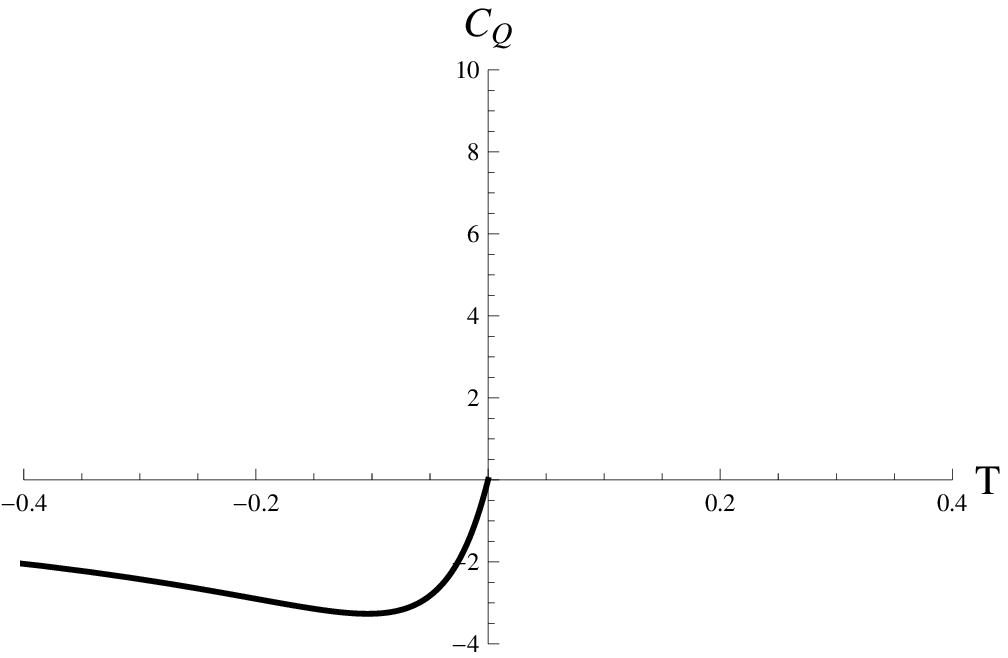}
  \end{minipage}
\caption{The variation of thermal entropy with temperature for
$Q=0.1,0.25,0.5,0.6$ respectively from top left in dS space time}
\end{figure}
It can be seen that upto $Q=0.5$ it shows a phase transition and
then after reaching $Q=0.6$, it no more shows any phase transition.
Also it is found that above this value no phase transition is
observed.\\ \\
The variation of the behavior of specific heat with $Q$ for the AdS
space time for the values $Q=0.1,0.25,0.3,0.4$ is shown in FIg $13$.
\begin{figure}[h]
    \begin{minipage}{\columnwidth}
  \includegraphics[width=9pc]{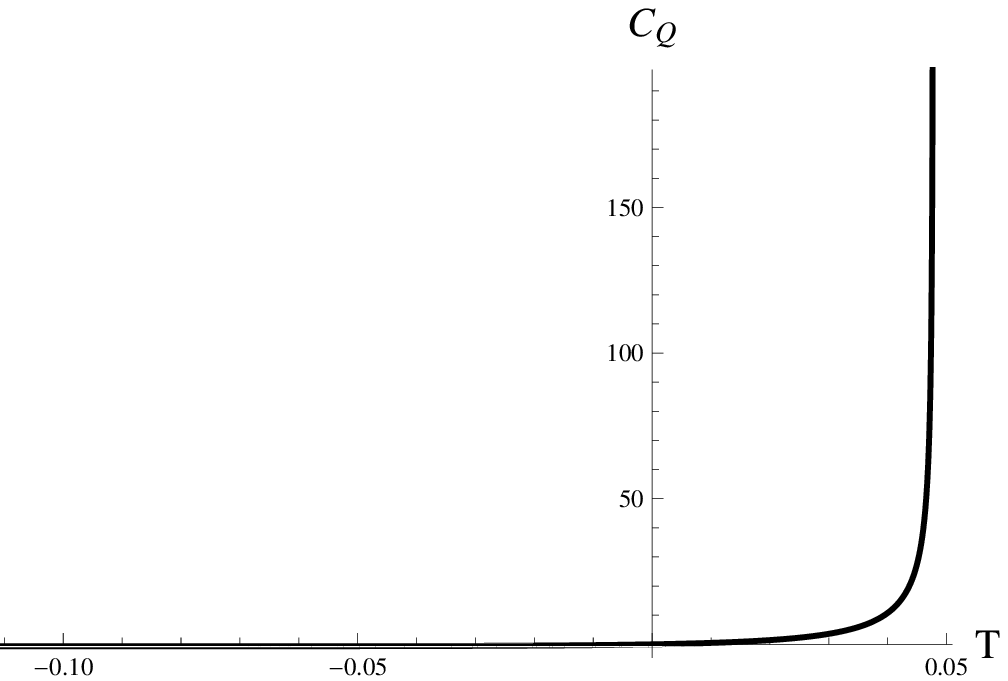}\hspace{0.5cm}\includegraphics[width=9pc]{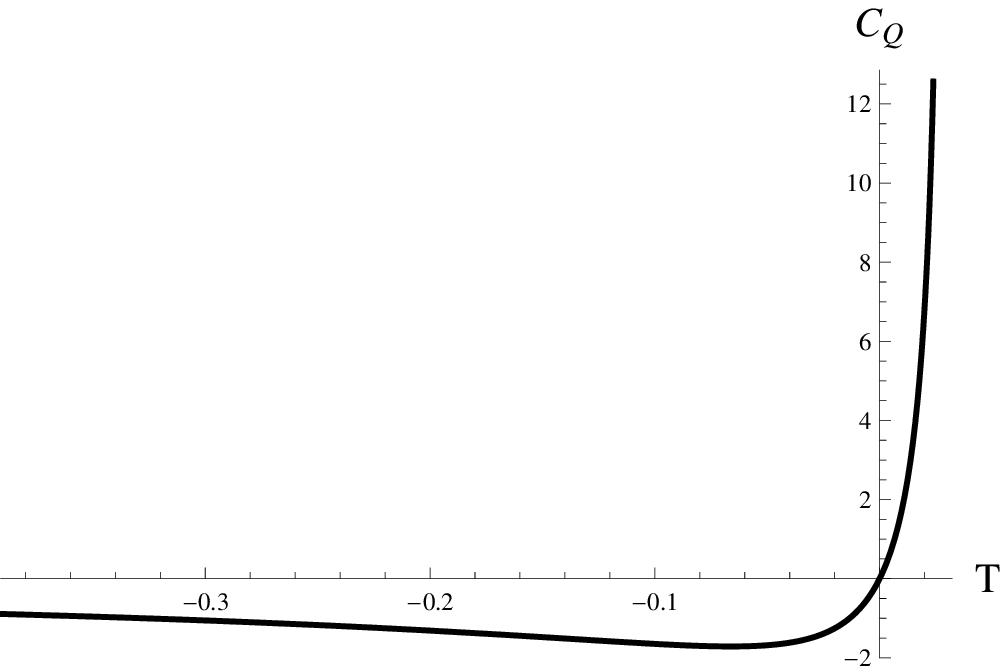}\\
  \includegraphics[width=9pc]{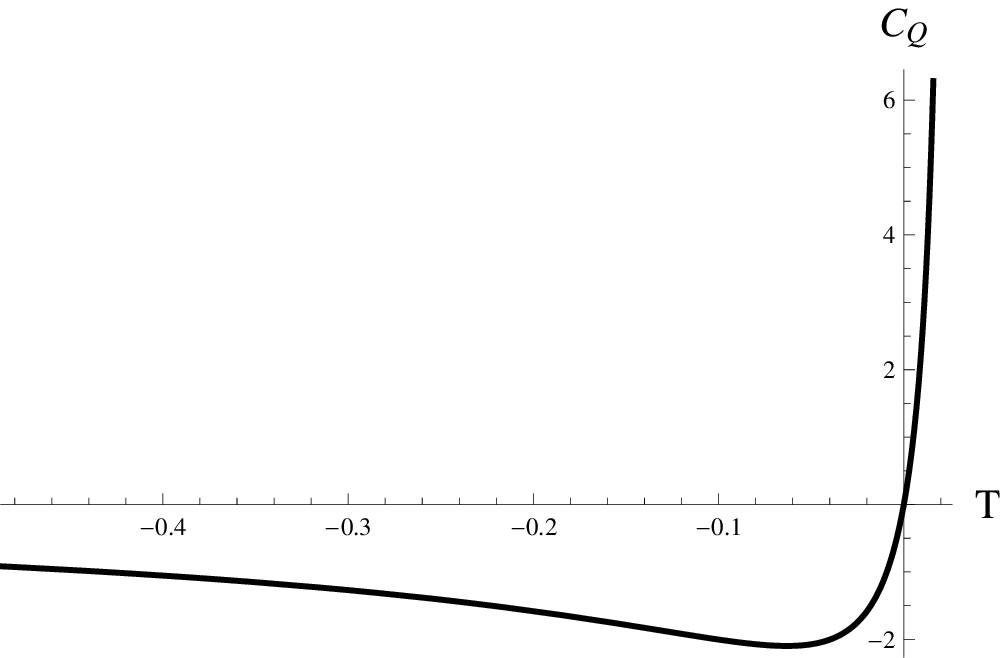}\hspace{0.5cm}\includegraphics[width=9pc]{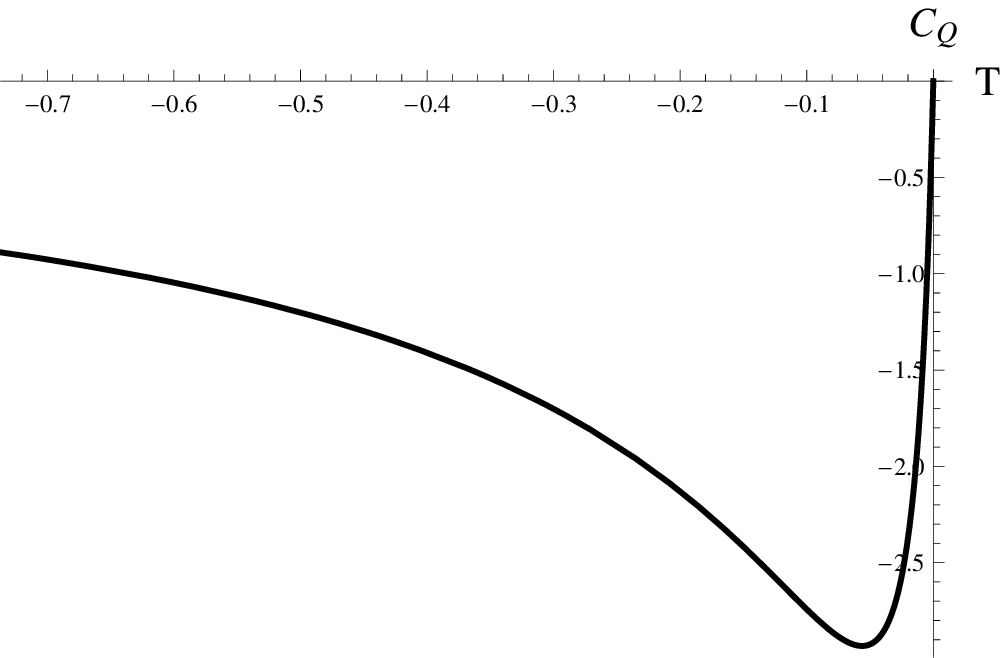}\\
  \end{minipage}
\caption{The variation of thermal entropy with temperature for
$Q=0.1,0.25,0.3,0.4$ respectively from top left in AdS space time}
\end{figure}
It can be seen that for $Q=0.1$ it does not show any phase
transition and upto $Q=0.3$ it shows a phase transition and then
after reaching $Q=0.4$, it no more shows any phase transition. Also
it is found that above this value no phase transition is observed.
From this it can also be concluded that AdS space time shows phase
transition only for a small range of $Q$ when compared with the dS
space time.

\section{Conclusion}
In this paper we have calculated the QNMs for a linearly charged BTZ
black hole in massive gravity. The values of the parameters are so
chosen that in the metric function, the massive parameter dominates.
It is found that in the de Sitter space time as the cosmological
constant $\Lambda$ is increased, the quasi normal frequencies varied
continuously and then after reaching a particular value of
$\Lambda(=0.1)$, their behavior is found to be abruptly changing
afterwards. This is shown in the $\omega_I-\omega_R$ plot where
there is a drastic change in the slope of the curve after a
particular value of $\Lambda$. This can be seen as a strong
indication of a possible phase transition occurring in the system.
When the massive parameter $m$ is increased, a similar behavior is
found but the $\Lambda$ at which the change of behavior of QNMs is
found to be shifted to a higher value$(\Lambda=0.28)$. Also, it can
be inferred that the variation of the massive parameter will only
alter the point at which the phase transition happens. For different
values of $Q$ the phase transition occurs for different values  of $\Lambda$.\\\\
The QNMs for an (Anti)de Sitter space time is also calculated and
the behavior of their quasi normal frequencies are analyzed. For
$Q=0.1$ the behavior of QNMs showed an inflection point but no phase
transition. However for $Q=0.25$ it showed a phase transition. Thus
it is seen that the phase transition behavior is found dependent on
$Q$ for the AdS case. It is also observed by studying the variation
of QNMs with $Q$ that AdS space time shows phase transition only for
certain limited ranges of $Q$ compared to the dS case.\\ \\
The thermodynamics of such black holes in the dS space is then
looked into. The behavior of specific heat showed phase transition
for the dS case for a wide range of $Q$ whereas for AdS space time
phase transition is shown only for a limited range of $Q$.
\section*{Acknowledgements} One of us (PP) would like to thank UGC,
New Delhi for financial support through the award of a Junior
Research Fellowship(JRF) during the period $2010$-$2013$. PP would
also like to acknowledge Govt. College, Chittur for allowing to
pursue her research. VCK would like to acknowledge Associateship of
IUCAA, Pune.


\begin{thebibliography}{}
\bibitem{eref1}Salvatore Capozziello and Mariafelicia De Laurentis, Phys. Rep.
           509,167(2011)(arXiv:1108.6266)
\bibitem{1}Timothy Clinton,Phys.Rep.513,1(2012)
\bibitem{2}Mattingly, D.,Liv.Rev.Rel.8,lrr-2005-5(2005)
\bibitem{3}Fierz, M. and Pauli, W.,Proc. R.Soc.Lond.Ser.A. 173, 211232(1939)
\bibitem{4}van Dam H.and Veltman M.J.G., Nucl.Phys.B, 22,397(11970)
\bibitem{5}Zakharov V. I.,JETP Lett.,12,312(1970)
\bibitem{6}Vainshtein, A. I.,Phys. Lett. B.39,393(1972)
\bibitem{7}Boulware,D.G.and Deser S., Phys. Rev. D6,3368(1972)
\bibitem{8}de Rham, C., Gabadadze, G. and Tolley, A.J.,Phys.Rev.Lett.106,231101(2011)
\bibitem{9}Volkov, M. S., Class. Quantum Grav. 30 184009 (2013)
\bibitem{10}Kodama H. and Arraut I,Prog.Theor.Exp.Phys.023E0(2014)(arXiv:1312.0370)
\bibitem{11}S. G. Ghosh, L.Tannukij and P. Wongjun, Eur. Phys. J. C. 76 (2016) 119.
\bibitem{12}P.Prasia and V.C.Kuriakose, Gen. Rel.Gravit.48,89(2016)
\bibitem{13}D.Vegh,CERN-PH-TH/2013-357(2013) [arXiv:1301.0537].
\bibitem{14}Hawking,S.W.and Page,D.N., Commun. Math. Phys. 87,577(1983)
\bibitem{15}P.C.W.Davies,Rep. Prog. Phys. 41,1313(1978)
\bibitem{16}Gross D.J.,Perry M.J. and Yaffe L.G., Phys. Rev.D25,330(1982)
\bibitem{17}S.Carlip arXiv 1410.1486
\bibitem{18}Brian P Dolan.,Class. Quantum Grav.28,125020(2011)
\bibitem{19}K.Ghaderi and B.Malakolkalami., Nucl. Phys.B,903,10(2016)
\bibitem{20}Jishnu Suresh, R. Tharanath, Nijo Varghese and V.C.
            Kuriakose, Eur. Phys. J C., 74, 2819(2014)
\bibitem{21}Fabio Capelaa and Peter G. Tinyakov., JHEP 1104:042(2011) (arXiv:1102.0479)
\bibitem{22}R.G. Cai, Y. P. Hu, Q. Y. Pan and Y. L. Zhang, Phys. Rev. D.91,024032(2015)
\bibitem{23}M. Banados, C. Teitelboim, and J. Zanelli, Phys. Rev. Lett.69,1849 (1992).
\bibitem{24}M. Banados, M. Henneaux,C. Teitelboim  and J. Zanelli, Phys. RevD.48,1506 (1992).
\bibitem{25}Norman Cruz and Samuel Lepe, Phys. Lett.B,593,235(2004)
\bibitem{26}Mariano Cadoni and Maurizio Melis, Found. Phys. 40,638(2010)
           (arXiv:0907.1559)
\bibitem{27}A. Chamblin, R. Emparan, C. Johnson, and R. Myers, Phys.Rev. D60,064018(1999)
\bibitem{28}M. Cadoni, M. Melis  and  M.R. Setare, Class. Quantum Grav. 25,195022(2008)
\bibitem{29}M.R. Setare  and  H. Adami, Phys. RevD, 91, 104039(2015)
\bibitem{30}Vishveswara, C.V., Nature 227, 936(1970)
\bibitem{31}S.Chandrasekhar and S. Detweiler, Proc. R. Soc. London A344,441(1975).
\bibitem{32}V. Cardoso and J. P. S. Lemos, Phys. Rev. D 63, 124015(2001)
\bibitem{33}Debaprasad Maity et al., Nucl. Phys. B839:526,(2010) (arXiv:0909.4051v2).
\bibitem{34}S. H. Hendi, B. Eslam Panah and S. Panahiyan, JHEP 05,029(2016)(arXiv:1604.00370v1)
\bibitem{35}Moss I. G. and Norman J. P., Class. Quant. Grav.19,2323(2002)
\bibitem{36}Ciftci H.,Hall R.L. and Saad, N.,Phys. Lett. A,340,388(2005)
\bibitem{37}Cho H.T.,Cornell, A.S., Jason, D., Huang, T.R.and Wade N.,
            Adv. Math. Phys.,doi:10.1555/2012/281705(2012)
\bibitem{38}Cho H.T.,Cornell, A.S.,Jason, D. and Wade N., Class. Quantum Grav.27,155004(2010)
\bibitem{39}J Xi, L Cao  and Y-P Hu, Phys. RevD.,91,124033(2015)
\end{thebibliography}
\end{document}